\documentclass[aip, amsmath, amssymb,
 reprint%
]{revtex4-2}

\usepackage{dcolumn}
\usepackage{bm}
\usepackage{physics}
\usepackage[version=4]{mhchem}

\allowdisplaybreaks 

\usepackage[utf8]{inputenc}
\usepackage[T1]{fontenc}
\usepackage{mathptmx}
\usepackage{etoolbox}

\usepackage{xcolor}

\usepackage{graphicx}
\usepackage{float}

\newcommand{\UPMF}{U_{\text{PMF}}}

\newcommand{\numb}[1]{#1^{\dagger}#1}

\makeatletter
\def\@email#1#2{%
 \endgroup
 \patchcmd{\titleblock@produce}
  {\frontmatter@RRAPformat}
  {\frontmatter@RRAPformat{\produce@RRAP{*#1\href{mailto:#2}{#2}}}\frontmatter@RRAPformat}
  {}{}
}%
\makeatother

\draft 

\begin{document}

\preprint{AIP/123-QED}

\title[Electronic Friction RPMD]{Electronic Friction Near Metal Surface: Incorporating Nuclear Quantum Effect with Ring Polymer Molecular Dynamics}
\author{Rui-Hao Bi}
\affiliation{Department of Chemistry, School of Science, Westlake University, Hangzhou, Zhejiang 310024, China}
\affiliation{Institute of Natural Sciences, Westlake Institute for Advanced Study, Hangzhou, Zhejiang 310024, China}

\author{Wenjie Dou}%
\affiliation{Department of Chemistry, School of Science, Westlake University, Hangzhou, Zhejiang 310024, China}
\affiliation{Institute of Natural Sciences, Westlake Institute for Advanced Study, Hangzhou, Zhejiang 310024, China}
\affiliation{Department of Physics, School of Science, Westlake University, Hangzhou, Zhejiang 310024, China}
\email{douwenjie@westlake.edu.cn} 

\date{\today}

\begin{abstract}
Molecular dynamics with electronic friction (MDEF) approach can describe nonadiabatic effects accurately at metal surfaces in the weak nonadiabatic limit. 
That being said, MDEF treats nuclear motion classically, such that the nuclear quantum effects are missing completely in the approach. 
To address this limitation, we combine electronic friction with Ring Polymer Molecular Dynamics (RPMD). 
In particular, we apply the averaged electronic friction from the metal surface to the centroid mode of the ring polymer. 
We benchmark our approach against quantum dynamics to show that electronic friction with RPMD (EF-RPMD) can capture zero-point energy as well as transition dynamics accurately. 
In addition, we show EF-RPMD can correctly predict the electronic transfer rate near metal surfaces in the tunneling limit as well as the barrier crossing limit. 
We expect our approach will be very useful to study nonadiabatic dynamics near metal surface when nuclear quantum effects become essential. 

\end{abstract}

\maketitle

\section{\label{sec:intro} Introduction}

Nonadiabatic dynamics at the metal-molecule interface have attracted significant attention in the chemical physics community \cite{Rev-HASSELBRINK2006, Rev-Wodtke2016}.
The breakdown of the Born-Oppenheimer approximation has been confirmed through a series of experiments on surface scattering events \cite{Vibrelax-Huang2000, HotElectronEmission-White2005}.
At the same time, the nonadiabatic effects also play important roles in a variety of fields such as chemisorption \cite{HAdsporption-Bunermann2015, AgSurfNODiss_Krger2016, AgSurfHClDiss-Geweke2020}, electrochemistry \cite{PCET-Lam2020, Surface-Lee2022}, heterogeneous catalysis \cite{NiMethane-Luo2016, H-Dorenkamp2018}, and molecular junctions \cite{Mjunction-Ke2021, CISS-Teh2022}.
To gain fundamental understandings to these processes in complex systems, theoretical simulations that accounts for the nonadiabatic energy and electron transfers become essential.
However, to account for the nonadiabatic effects in molecular dynamics at metal surface is very challenging, where the Born-Oppenheimer approximation is not valid any more and the coupled dynamics of nuclear and electronic degrees of freedoms become relevant \cite{Rev-HASSELBRINK2006, Rev-Wodtke2016}. 

Among the theoretical approaches to address the breakdown of the Born-Oppenheimer approximation at metal surfaces, classical trajectories based methods are probably the most useful ones. Despite that 
there exists exact and approximate quantum treatment on the coupled dynamics \cite{MCTDH-Thoss2007, QMC-Mhlbacher2008, HQME-Schinabeck2016, HQME-Xu2019}, the computational costs of these methods are often very demanding to be applied to large/realistic chemical systems.
Due to the fact that many nuclear DoFs are involved in realistic systems, classical treatment on the nuclear dynamics with quantum tretament on the electronic DoFs are often more practical, resulting in so called mixed quantum-classical methods \cite{REV_MQC-CrespoOtero2018}.
Many recent developed methods belong to this family, such as Independent Electronic Surface Hopping \cite{JCP-Shenvi2009, Science-Shenvi2009}, Classical Master Equations \cite{QCMEalg-D2015, QCMEderive-D2015, MultiLevel-Dou2017}, and Molecular Dynamics with Electronic Friction (MDEF) \cite{HeadGordonFriction-1995, EF1L-D2015, EFML-D2016, EFPRL-D2017, EFREV-D2018}. In particular \cite{EFPRL-D2017, EFREV-D2018}, the MDEF method is probably the most straightforward one, where one runs Langevin dynamics for the nuclei and all electrons DoFs give rise to frictional forces and random forces. MDEF method has been implemented into \textit{ab initio} electronic structure calculations and has been proven to be useful to predict energy relaxation near metal surfaces \cite{AbinitoFriction-BlancoRey2014, AbinitoFriction-Maurer2016, maurer2017ModeSpecificElectronic, zhang2019HotelectronEffectsReactive}.

That being said, MDEF and essentially the other classical trajectories based methods (IESH, CME) fail  when nuclear quantum effects are important. 
The nuclear quantum effects (NQEs) including the zero-point energy and tunneling behaviors become relevant when dealing with high frequency motions of light atomic nucleus such as hydrogen at low temperatures \cite{NQE-Markland2018}.
As a result, the mixed quantum-classical approaches cannot be directly applied to study the "non-classical" nucleus, and additional treatments are usually required \cite{REV_MQC-CrespoOtero2018, RPSH-Shushkov2012, SH-infiniswap_Lu2018, isomorphic-Tao2018, Mapping-Richardson2013, MVRPMD-Ananth2013, IB-Zhao2023}.
Incorporating NQEs in molecular dynamics has been studied extensively in solution or gas phase. Except for a few recent studies, little attention is being paid on NQEs near a metal surface 
For instance, De \textit{et al.} have shown that one can incorporate NQEs via a flavor of Independent Electronic Surface Hopping (IESH-D) \cite{IESH-D_Amber2023}.
Meanwhile, 
Litman \textit{et al.} have proposed an instanton rate formalism based on \textit{ab initio} MEDF to include NQEs \cite{TheoryEF_Rossi2022}. 
Using this method, they can calculate the tunneling rate for model system and hydrogen diffusion on metal surface \cite{NumEF_Rossi2022}.
Despite these efforts, efficient method that predict accurate transient dynamics in the NQE regime is not readily available.

In this work, we combine the RPMD method with MDEF to include the NQEs. We show that one can add straightforward averaged electronic friction (and random force) to the centroid mode of the ring polymer to capture electronic weak-nonadiabatic effects near metal surfaces.
We validate our method on steady state distribution and/or population as well as on transient dynamics with nearly exact quantum dynamics, where we reach to perfect agreement. Moreover, we demonstrate that our method can predict the electron transfer rate that agrees with the exact tunneling rate at low temperature and reproduce the Marcus barrier crossing rate at high temperature. 
Given the simplicity and accuracy of the method, we expect EF-PRMD will be very useful in studing nonadibatic dynamics near metal surface with NQEs. 

The paper is organized as follows: Section~\ref{sec:theory} introduces the standard model for an molecule on a metal interface and the electronic friction theory associated with such a model. We then explain our rationale for extending classical trajectory methods derived from electronic friction theory with RPMD. In Section~\ref{sec:numeric}, we present several numerical tests on the RPMD methods at low temperatures, validated against results obtained using the Quantum Master Equation (QME). In Section~\ref{sec:conclusion}, we conclude. 

\section{\label{sec:theory} Theory}

\subsection{\label{subsec:adsorption} The Chemisorption Model}
To model the dynamics of a molecule on a metal surface, we employ a Newns-Anderson Hamiltonian \cite{Adsorption-NEWNS1969},
which describes the a single orbital of the molecule coupled to a continuum of electronic states from the metal.
Specifically, we consider:
\begin{gather}
		H   = H_s + H_b + H_c,                           \label{eqn:totH}\\
		H_\text{s} = \frac{P^2}{2M} + E(x) d^{\dagger} d +%
                      V_0 (x),       \label{eqn:Hs}\\
		H_\text{b} = \sum_k (\epsilon_k - \mu)c^{\dagger}_k c_k,       \label{eqn:Hb}\\
		H_\text{c}= \sum_k V_k( c_k^{\dagger} d +%
               d^{\dagger} c_k),                         \label{eqn:Hc}
\end{gather}
Here, $H_\text{s}$ describes the molecule that includes a electronic level with corresponding creation (annihilation) operator $d^{\dagger}$ ($d$) and a nuclear degree of freedom (DOF) with corresponding position and momentum operators $x$ and $P$. $E(x)$ is the on-site energy of the molecular orbital and $V_0(x)$ is the external nuclear potential.
$H_\text{b}$ represents a bath consisting of a continuum of electrons with corresponding creation (annihilation) operator $c_k^{\dagger}$ ($c_k$). $\mu$ denotes chemical potential and $T$ denotes temperature. 
$H_\text{c}$ represents the interaction between the molecular orbital and the metal. We can define the hybridization function $\Gamma$ to describe the strength of the couplings: 
\begin{gather}
\Gamma = 2\pi \sum_k \abs{V_k}^2 %
                 \delta(\epsilon_k - \epsilon).          \label{eqn:WBA}
\end{gather}
To further simplify the interaction, we will apply the wide-band limit, such that the hybridization function $\Gamma$ is a constant, which does not depend on energy $\epsilon$ nor position $x$.  

Without loss of generality, we assume the nuclear potential is a harmonic oscillator with frequency $\omega$. 
We further assume that the on-site energy $E(x) = \sqrt{2M\omega/\hbar} g  x + E_d$, where $g$ denotes the strength of linear electron-phonon coupling, and $E_d$ the position-independent orbital energy. 
With these simplifications, we have two diabatic PESs for the neutral and charged state: 
\begin{equation*}
    V_0(x) = \frac{1}{2} M \omega^2 x^2, \quad V_1(x) = V_0(x) + \sqrt{\frac{2M\omega}{\hbar}} g x + E_d.
\end{equation*}
Note that the charged state parabola $V_1$ has its equilibrium position shifted by $\frac{\sqrt{2}g}{\omega} (M\hbar\omega)^{-1/2}$, and its equilibrium energy lowered by $\tilde{E}_d = E_d - g^2 / \hbar \omega$. 


Overall, we have introduced a concrete model for metal surface electron transfer. 
In the following, we will briefly mention how to study this model in the classic temperature regime using molecular dynamics (MDEF). 

\subsection{\label{subsec:friction} Electronic Friction and Molecular Dynamics}
In the area of chemisorption and electrochemistry, electron friction has been utilized to incorporate weak nonadiabatic effects, particularly for the dynamics of charge transfer between a molecule and a metal surface \cite{HeadGordonFriction-1995, Rev-HASSELBRINK2006, lu2012CurrentinducedAtomicDynamics, Rev-Wodtke2016, EFPRL-D2017}.
Through electron-phonon couplings, molecules can dissipate its vibration energy into the electronic excitations, which results in electron-hole pairs (EHPs) in the metal.
The creation and re-combination of EHPs give rise to a frictional force as well as a fluctuating force onto the molecule \cite{HeadGordonFriction-1995, Rev-Wodtke2016, hertl2021RandomForceMolecular}.
Electronic friction described above is the first order corrections to the Born-Oppenheimer approximation \cite{EFREV-D2018}. 
Such correction can be readily included in molecular dynamics with friction and random force as described by Langevin equation \cite{EF1L-D2015, EFML-D2016, EFPRL-D2017, EFREV-D2018}. We now briefly introduce the Electronic Friction Langevin Dynamics method before we address the nuclear quantum effects. 

When the molecule interacts with metal strongly with electrons exchanging rapidly between them (i.e., when $\Gamma$ is not too small), we can map the dynamics of the total system into a Fokker-Planck equation \cite{EF1L-D2015}:
\begin{equation}
    \label{eqn:FP}
	\begin{aligned}
		\pdv{A(x, P, t)}{t} = & -\frac{P}{M} \pdv{A(x, P, t)}{x}   %
		-\bar{F}\pdv{A(x, P, t)}{p} +                          \\
		& \gamma_\text{e} \pdv{P} [\frac{P}{M} A(x, P, t)] %
		+ D\pdv[2]{A(x, P, t)}{P}.
	\end{aligned}
\end{equation}
Here $A(x, P)$ denotes the total phase space density for the nuclei. 
The first two terms on the right hand side of the above equation denote the classical motion on the potential of mean force. 
Whereas the last two terms denote dissipation and fluctuation from the electronic motion. 
Within this context, $\bar{F}$, $\gamma_\text{e}$, and $D$ denote the mean force, friction, and correlation function of the random force, correspondingly defined by the following equations:
\begin{gather}
    \bar{F}(x)  = -\pdv{E}{x} \sigma_\text{eq}(x) - \pdv{V_0}{x}, \quad \sigma_\text{eq}(x) = f(E(x)) \label{eqn:PMF}\\ %
    \gamma_\text{e}(x) = \frac{\hbar}{\Gamma} \left(\pdv{E}{x}\right)^2 \frac{\sigma_\text{eq} (1 -\sigma_\text{eq})}{k_\text{B}T}, \\%
    D(x) = k_\text{B} T \gamma_\text{e}(x) \label{eqn:fluc-diss}.
\end{gather}
We note all these quantities depends on $\sigma_\text{eq}(x)$, which represents the instantaneous equilibrium population of the molecular level at position $x$, i.e., $\sigma_\text{eq}(x) = \expval{d^{\dagger}d}(x) = f(E(x))$. 
Here, $f(\epsilon)$ denotes the Fermi-Dirac distribution function, $f(\epsilon) = (1 + e^{\beta (E-\mu)})^{-1}$ (where $\beta = 1/k_\text{B} T$ denotes the inverse temperature).

In practice, the dynamics encoded in the Fokker-Planck equation (Eq.~\ref{eqn:FP}) can be easily solved through molecular dynamics simulations. 
Specifically, we numerically integrate a swarm of trajectories according to the following Lagevin equation
\begin{equation}
    \label{eqn:langevin}
    \dot{x} = \frac{P}{M}, \quad \dot{P} = \bar{F} - \gamma_\text{e}(x) \frac{P}{M} + \delta F(t),
\end{equation}
Here $\delta F(t)$ is a fluctuating random force that satisfies a Markovian correlation function as follows: 
\begin{equation} \label{eq:correlation_function}
    \expval{\delta F(t) \delta F(t^{\prime})} = 2 D(x) \delta(t - t^{\prime}).
\end{equation}
The above correlation function corresponds to Gaussian white noises, which can be readily generated to simulate the random force. 
Overall, the swarm of Langevin dynamics trajectories represent the evolution of the phase space density, which is equivalent to the Fokker-Planck equation (Eq.~\ref{eqn:FP}).

That being said, we need to acknowledge the Langevin dynamics described above is only applicable in the weak nonadiabatic regime. Such limitation can be understood from a mean field perspective \cite{EFPRL-D2017, EFREV-D2018}. 
Specifically, when electronic relaxation is significantly faster than the nuclear motion, the first order expansion of the electronic density with respect to momentum is valid. 
Moreover, this allow us to assume the random force correlation function is Markovian.
In the context of our chemisorption model, the Langevin dynamics is only valid when $\Gamma > \hbar \omega$. 
In Appendix~\ref{app:breakdown}, we demonstrate such insight lead to the conclusion that MDEF corresponds to Ehrefenst dynamics with random forces.

Overall, we have briefly introduced the MDEF formalism for studying the transient dynamics near metal surfaces with the weak nonadiabatic effects.
Despite the its success in the classical regime, the NQEs are completely missing out.
It is evident, when $kT < \hbar \omega$, the Newtonian dynamics will deviate from the true quantum dynamics for two reasons:
1) The phase space probability from the Newtonian dynamics deviates from the quantum distribution in low temperatures. 
2) a "local" electronic friction $\gamma_\text{e}(x)$ calculated from a definite position $x$ is not applicable in the quantum limit.
To address these limitations, we propose a straightforward ring-polymer molecular dynamics (RPMD) extension of the MDEF approach in the following section.

\subsection{\label{subsec:RPMD} Electronic Friction for a Ring Polymer}

\begin{figure}[htbp]
	\centering
    \includegraphics[width=.48\textwidth]{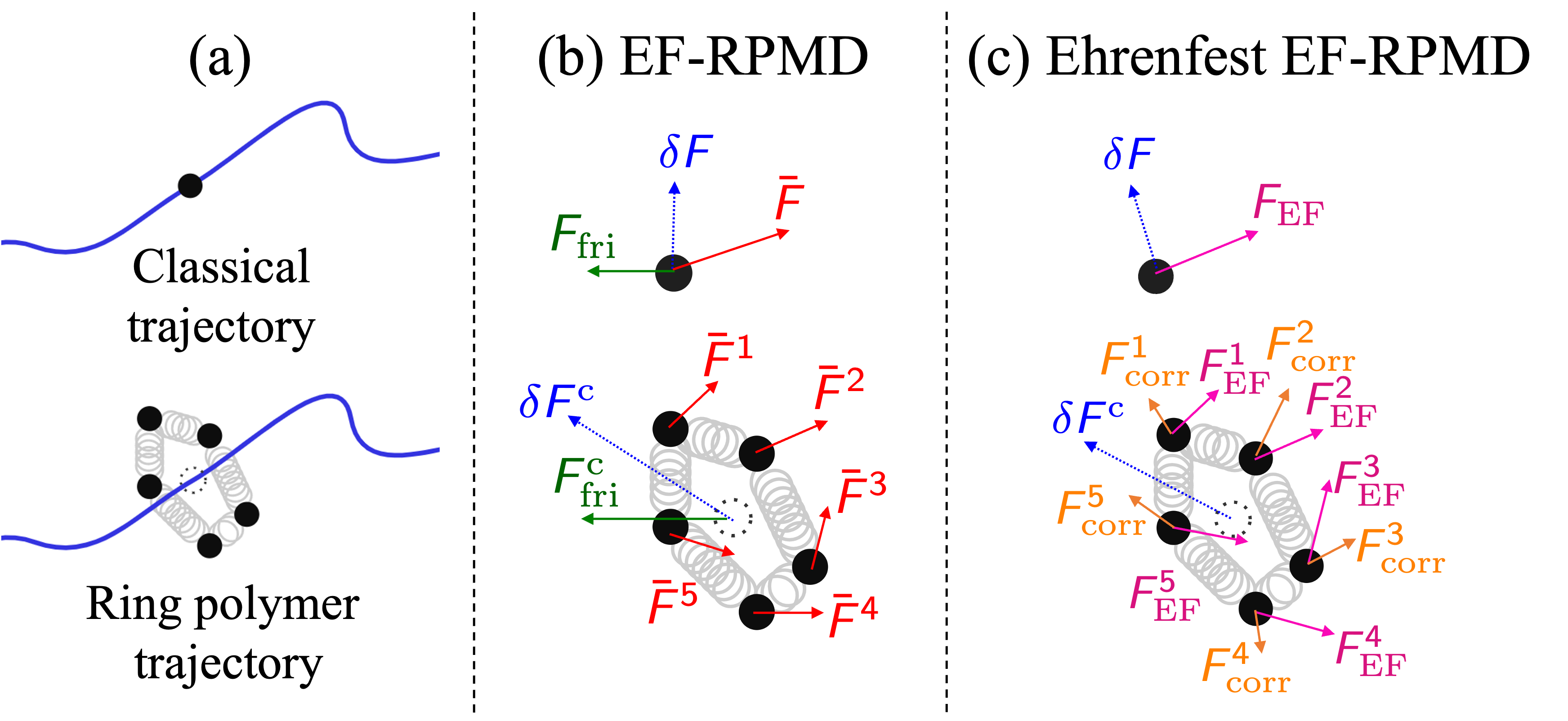}
	\caption{\label{fig:theme}
        Schematic representation comparing a classical nucleus with a ring polymer in the context of electronic frictional forces. 
        In Panel (a), the ring polymer's centroid is depicted as an analogue of a classical nucleus. 
        Panels (b) and (c) highlight the distinctions in electronic frictional forces between a classical nucleus and a ring polymer for EF-RPMD, along with its mean-field counterpart--Ehrenfest EF-RPMD, respectively.}
\end{figure}

In FIG.~\ref{fig:theme}, we sketch a simple scheme to include nuclear quantum effects with electronic friction, henceforth denoted as EF-RPMD. To be more explicit, in the quantum regime, the a classical particle is replaced by the multiple replicas of the particle, hence a ring polymer. Each bead in the ring polymer feels a local mean force $\bar{F}$. The electronic friction and random force are only added onto the centroid mode of the ring polymer. The electronic friction and random force, however, are the averaged friction and random force for each bead at different positions. This simple scheme for adding frictional effects from electron motion in RPMD is shown in FIG.~\ref{fig:theme}(b). 

In FIG.~\ref{fig:theme}(c), we sketch a simple scheme to include nuclear quantum effects for the Ehrenfest dynamics. Because of the connection between the Ehrenfest dynamics and the electronic friction molecular dynamics in the weak nonadiabatic limit, we make sure that the Ehrenfest dynamics reproduce EF-RPMD dynamics in this limit. Details of including nuclear quantum effects in Ehrenfest dynamics are shown in Appendix~\ref{app:breakdown}. Here in the main text, we focus on the EF-RPMD dynamics. 

The system Hamiltonian $H_s^{N}$ with $N$ ring polymer beads can be written down straightforwardly as follows: 
\begin{equation}
	\label{eqn:Hs_path}
	\begin{gathered}
		H_s^{N}  = H_0^{N} + E^{N} + V_0^{N}, \\
        H_0^{N} = \sum_{i=1}^{N} \left[ \frac{P_i^2}{2M} + %
			\frac{1}{2}M \omega_N^2(x_i - x_{i+1})^2 \right], \\%
        E^{N} = \sum_{i=1}^{N} d^{\dagger} d E(x_i), \quad%
		V_0^{N} = \sum_{i=1}^{N} \frac{1}{2} M \omega^2 x_i^2.
	\end{gathered}
\end{equation}
Here, $H_0^{N}$, $E^{N}$, and $V_0^{N}$ represent the "free" ring polymer, on-site energy of the molecular orbital for the ring polymer, and nuclear potential components of the Hamiltonian, respectively. 
$x_i$ and $P_i$ indicate the position and momentum of the $i$-th bead. The bead index $i$ is subjected to periodic boundary conditions, such that $i = N + i$. 
$\omega_N \equiv k_\text{B} T N / \hbar$ denotes the spring constant of the spring that connects neighboring beads.

When the electrons are moving fast enough, the electronic part of the system Hamiltonian can be integrated out, such that we can use the averaged electronic population to the electronic degree of freedom $d^{\dagger}d$.
In other words, the electronic DoFs of each ring polymer bead $i$ reach to the instantaneous equilibrium $\sigma_\text{eq}(x_i) = f(E(x_i))$. As a result, bead $i$ feels a mean force $\bar{F}^{(i)}$ defined in the following equation,
\begin{equation}
    \label{eqn:EF-RPMDmeanF}
    \bar{F}^{(i)} = -\pdv{E}{x_i} \sigma_\text{eq}(x_i) -\pdv{V_0^{N}}{x_i}.
\end{equation}

We now turn to the question of how to include nonadiabatic effects known as "electronic friction" in a ring polymer. Naively, one can include electronic friction and random force on each bead at different positions. 
However, this scheme will only introduce internal friction and random force within the ring polymer---where the random force on each ring polymer bead can cancel each other, resulting in the absence of an external random force on the entire ring.
As a result, the second fluctuation and dissipation theorems does not satisfied. With this intuition in mind, we add the friction and random force from the electronic bath on the centroid mode. 
Specifically, the centroid mode of the ring polymer feels an averaged electronic friction with coefficients:
\begin{gather}
    \gamma_\text{e}^\text{c} = \frac{1}{N} \sum_{i=1}^{N} \gamma_\text{e} (x_i), %
\end{gather}
Here, the centroid mode friction $\gamma_\text{e}^\text{c}$ is computed by averaging over beads (the "quantum" expectation value). 
Furthermore, centroid mode also feels a random force that satisfies the second fluctuation dissipation theorem,
\begin{equation}
    D^\text{c} \equiv \expval{\delta F^\text{c}(t) \delta F^\text{c}(t^{\prime})} = k_\text{B} T \gamma_\text{e}^\text{c}  \delta(t-t^{\prime}) \label{eqn:fluc-diss-RP}.
\end{equation}
We have applied the Markovian approximation which is valid in the weak nonadiabatic limit. 
Finally, the Langevin equations of motion for each bead can be written as:
\begin{equation}
    \label{eqn:centroid_mode_eom}
    \dot{P}_i  = \bar{F}^{(i)} -\gamma_\text{e}^\text{c} \frac{1}{N} \sum_j P_j + \delta F^\text{c}.
\end{equation}
Here, note that each ring polymer bead feels the same electronic friction and random force, which is effectively adding a collective averaged friction on the centroid. 

Thus far, we have outlined how to include nuclear quantum effects in electronic friction model within the framework of RPMD method. Detailed distribution of the method and the numerical algorithms can be found in the Appendix~\ref{app:integrator}. 
In the Appendix~\ref{app:breakdown}, we have also described how to include quantum effects in the Ehrenfest dynamics near metal surface using RPMD. 
Note that our methods are not rigorously derived from first principle. 
That being said, in the following section, we will benchmark our methods again nearly exact quantum treatment to valid the EF-RPMD. 
We will show that EF-RPMD not only captures the equilibrium distribution perfectly, but also predicts dynamics and electron transfer rate nearly exactly.

\section{\label{sec:numeric} Numerical Tests For RPMD With Friction}

In this section, we will validate the EF-RPMD and Ehrenfest EF-RPMD methods by comparing with accurate dynamics from the quantum master equation (QME).
In ref.~\cite{QCMEderive-D2015}, we have derived the QME for Hamiltonian $H$ (Eq.~\ref{eqn:totH}) at the level of Redfield theory, which is nearly exact when the system-bath coupling is small. 
Instead of treating nuclear degrees of freedom (DOFs) classically as variables ($x$ and $P$), the QME method uses quantized phonon states and corresponding operators to handle these DOFs.
The finite temperature dynamics can be obtained by integrating the Liouville-von Neumann equations for the density matrix expanded in phonon basis.
The dynamics can be converged at arbitrary temperatures by increasing the number of phonon states used in the simulation.

\subsection{\label{subsec:ZPM} Equilibrium Nuclear Distributions and Zero-point Motion}

\begin{center}
\begin{figure}[htbp] 
	\includegraphics[width=.95\textwidth]{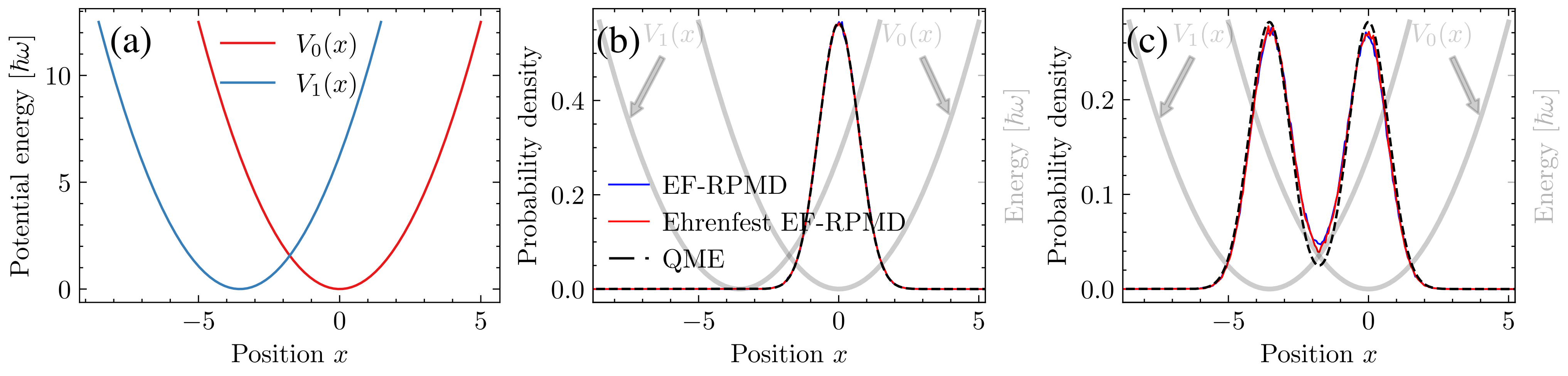}
	\caption{\label{fig:distribution}(a) The diabatic PESs $V_0(x)$ and $V_1(x)$: $\hbar \omega=0.003, g=0.0075, E_d=g^2/\hbar \omega$, and $\mu=0$. (b) and (c) are the initial and final position distributions, respectively, at $k_\text{B}T=0.1 \hbar\omega$. 
    Distributions from the QME were computed from the wavefunction of the initial and final states. 
    Distributions from the RPMD methods were obtained from the initial and final snapshots of 10,000 ring polymer trajectories.}
\end{figure}    
\end{center}

In FIG.~\ref{fig:distribution}, we plot the equilibrium nuclear distribution from EF-RPMD dynamics and Ehrenfest EF-RPMD dynamics. As shown in FIG.~\ref{fig:distribution}(a), we consider the PESs $V_0(x)$ and $V_1(x)$ for the symmetric case, where $\tilde{E}_d = 0$. 
Here, we note the barrier height for FIG.~\ref{fig:distribution} (a) is approximately $1.5 \hbar \omega$, which is relatively a shallow barrier. In the low temperature limit ($k_\text{B} T \ll \hbar \omega$), however, the barrier crossing process for the nuclei is classically forbidden. The zero-point energy and quantum tunneling are dominated in this limit.

\onecolumngrid
\twocolumngrid

We now show that the zero-point energy and quantum tunneling effects can be captured in the EF-RPMD methods. In FIG.~\ref{fig:distribution} (b), we show the initial nuclear position distribution from RPMD as well as the quantum master equation. The initial ring-polymer position distribution can be obtained from Monte Carlo samplings. Here $k_\text{B} T = 0.1 \hbar \omega$ is way below the barrier and the zero point energy (ZPE). Note that ring-polymer position distribution agrees with the quantum distribution, meaning that the delocalized effects in the low temperature is captured in RPMD. We then propagate the dynamics from Quantum master equation as well as EF-RPMD and Ehrenfest EF-RPMD. In FIG.~\ref{fig:distribution} (c), we plot the final nuclear distribution from these methods. Note that the equilibrium distribution of the ring polymer position agree perfectly with the quantum results. This agreement shows that the RPMD can capture the nuclear quantum effects and reach to the correct detailed balance.

\subsection{\label{subsec:ET} Electron Transfer for Symmetric PESs}
In the above subsection, we have shown that the RPMD methods can capture the nuclear quantum effects at steady state. In this subsection, we further show that the dynamics from EF-RPMD methods are correct as benchmarked against nearly exact quantum dynamics. In particular, we are interested in the electronic population dynamics in the molecule. 

\begin{figure}[htbp]
    \centering
    \includegraphics[width=0.48\textwidth]{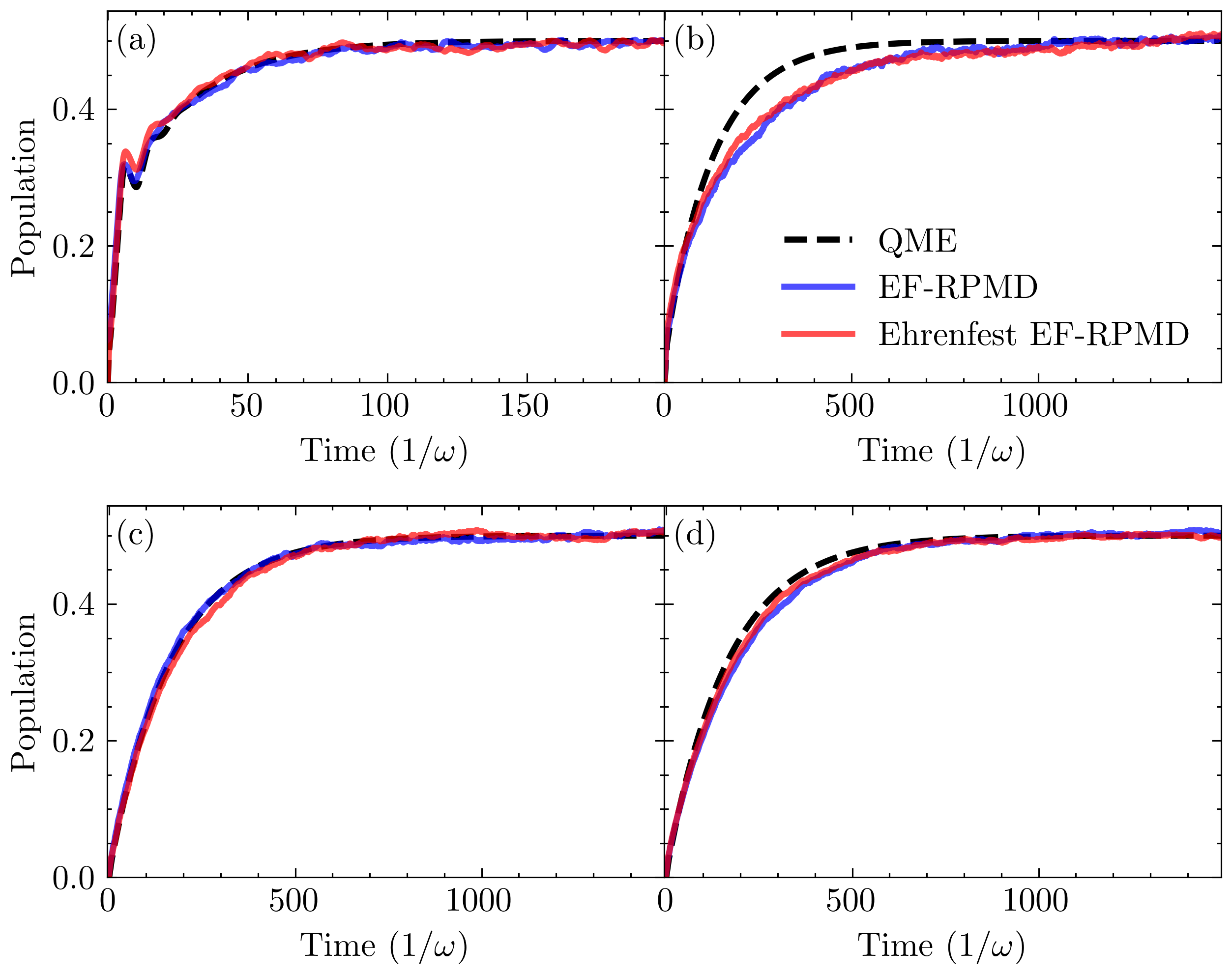}
    \caption{\label{fig:population}
    Electronic population of the single molecular level as a function of time for at different temperatures. 
    Panel (a-d) corresponds to $k_\text{B} T / \hbar\omega = 1, 0.333, 0.1, 0.033$, respectively.
    Black dashed-line represents the QME dynamics. Blue and red solid line represents the EF-RPMD and Ehrenfest EF-RPMD, respectively. 
    The RPMD population profiles were averaged from 10,000 trajectories, and the corresponding ring polymer size for panel (a-d) were 10, 40, 60, and 150 beads.
    Hybridization function $\Gamma=0.01 \approx 3 \hbar \omega$, i.e., weak non-adiabatic regime.
    All other simulation parameters were identical to those in FIG.~\ref{fig:distribution} (a).}
\end{figure}

In FIG.~\ref{fig:population}, we plot the electronic population of the molecular level at different temperatures for the symmetric PESs case (FIG.~\ref{fig:distribution} (a)). 
Remarkably, we find a nearly exact agreement between the population predicted by molecular dynamics and the QME dynamics across a broad range of temperature, from $2$ to $0.01$ $\hbar \omega$. 
Note that the RPMD is mostly being used to study steady state properties, whereas extending the RPMD for transient dynamics is not rigorously validated. 
Here near a metal surface, we are surprised to see that RPMD not only predicts the correct steady state population for the electron but also the transient dynamics.
Overall, our RPMD methods connect the high temperature and lower temperature limits, with the cost of increasing number of beads as the temperature decreases.

Note also that the EF-RPMD dynamics agree with its Ehrenfest counterpart well in FIG.~\ref{fig:population}. 
This is due to the fact that we are in the weak nonadiabatic regime ($\Gamma > \hbar\omega$, where the Ehrenfest dynamics can be mapped onto Markovian frictional and random force). 
That being said, in the strong nonadiabatic limit ($\hbar\omega > \Gamma$), EF-RPMD and Ehrenfest EF-RPMD should not agree with each other. 
We demonstrate the deviation between EF-RPMD and Ehrenfest EF-RPMD when $\hbar\omega > \Gamma$ in FIG.~\ref{fig:population-nonMark} (see Appendix.~\ref{app:breakdown}).

\begin{figure}[htbp]
    \centering
    \includegraphics[width=0.35\textwidth]{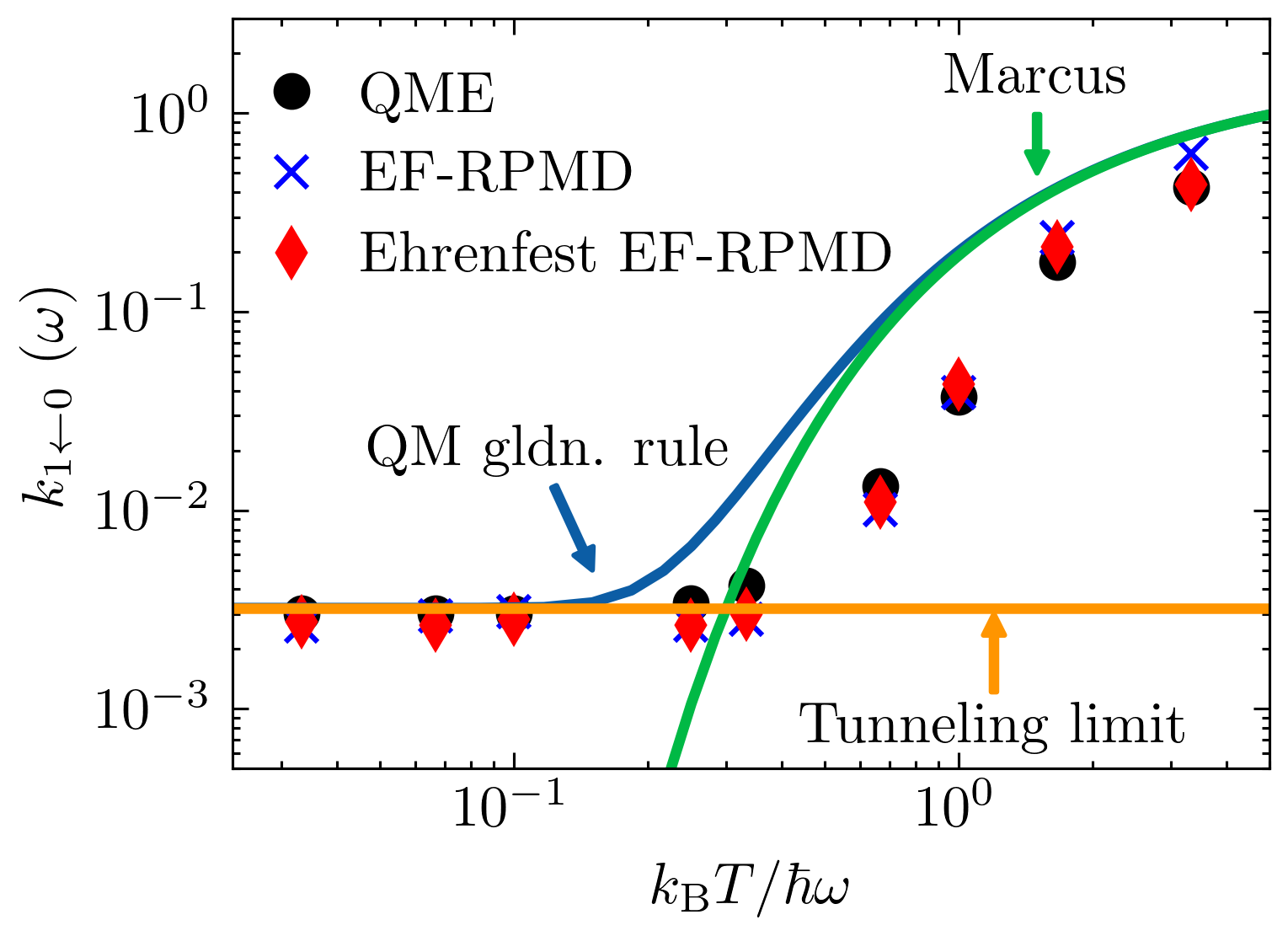}
    \caption{\label{fig:rate}
    Forward electron transfer rate $k_{1\gets0}$ as a function of temperatures. 
    $\Gamma=0.01 \gtrsim 3 \hbar \omega$, and all other parameters were identical to those in FIG.~\ref{fig:distribution} (a).
    Each scatter point for both RPLD and RPED was computed from 10,000 trajectories. The continuous lines denote various analytical results in different regimes, where dark-blue for quantum mechanical golden-rule rates, green for Marcus (barrier crossing) rates, and orange for low-temperature tunneling limits.} 
\end{figure}

In addition to the electronic dynamics, we can further compute the electron transfer rate by fitting these population dynamics exponentially. 
Specifically, we note that the electronic population $N(t)$ follows first-order kinetics, where the total electron transfer rate $k_\text{t}$ can be represented by the sum of the forward and backward electron transfer rates, denoted as $k_{1 \gets 0}$ and $k_{0 \gets 1}$ respectively. 
This relationship can be described by the exponential expression $N(t) = N_\text{eq} (1 - \exp(-k_\text{t} t))$, where $N_\text{eq}$ represents the equilibrium population \cite{QCMEderive-D2015}. For the symmetric PESs, we have $N_\text{eq}=0.5$. 
We can then quantify $k_{1\gets0}$ and $k_{0\gets1}$ with the detailed balance condition,
\begin{equation}
    \label{eqn:detailed_balance}
    k_{1\gets0} = e^{\beta \tilde{E}_d} k_{0\gets1},
\end{equation}
Here, $\tilde{E}_d$ is the renormalized energy for the molecular level.  

In FIG.~\ref{fig:rate}, we plot the forward rate $k_{1\gets0}$ as a function of temperature. 
As expected, both RPMD methods accurately reproduce the QME rates, further affirming the ability of RPMD for capturing correct dynamics as well as steady state population. In the high temperature limit, QME as well as the two RPMD methods reproduce the Marcus rate.  
Surprisingly, when comparing the rate estimated from the thermally averaged golden rule (as detailed in Appendix \ref{app:rate}), both EF-RPMD and Ehrenfest EF-RPMD predict the correct trend as temperatures decrease.
Specifically, the rates decrease until reaching a non-vanishing constant that remains independent of temperature.
As shown in Appendix.~\ref{app:rate}, this constant rate represents the tunneling limit of the golden-rule rate for electron transfer. FIG.~\ref{fig:rate} shows our numerical results agree with the quantum mechanical golden rule value quantitatively. 
This is strong evidence that our RPMD methods can capture the tunneling effect of electron transfer which dominates when temperatures are low. 
Hence, we conclude that the EF-RPMD and Ehrenfest EF-RPMD dynamics enables efficient exploration of both high-temperature barrier crossing and low-temperature nuclear tunneling regimes in electron transfer studies, necessitating only an increase in the number of beads.

\subsection{\label{subsec:detailedbalance} Electron Transfer for Asymmetric PESs}
In the previous subsection, we have demonstrated that our RPMD methods can perfectly describe both the steady state and dynamics of electron transfer processes for the symmetric PESs case. In this subsection, we show that for the asymmetric PESs, i.e. $\tilde{E}_d \neq 0$, the RPMD methods can deviate from the quantum results when the temperature is very low.

\begin{center}
\begin{figure}[htbp] 
	\includegraphics[width=.95\textwidth]{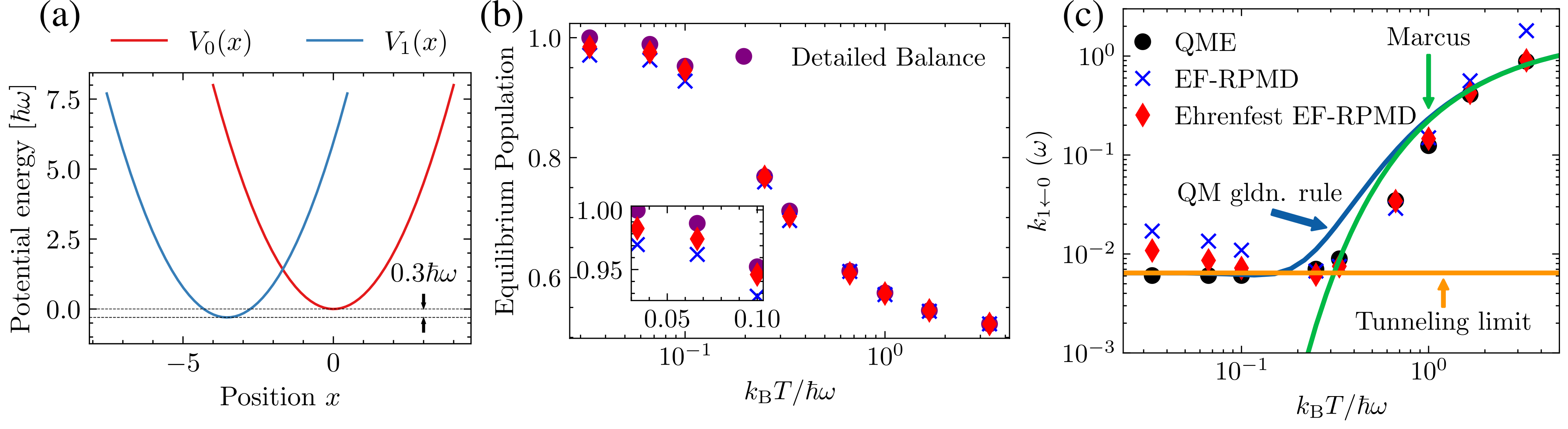}
	\caption{\label{fig:detailedbalance}
    (a) The diabatic PESs $V_0(x)$ and $V_1(x)$. Except $E_d=g^2/\hbar \omega + 0.3\hbar$, other parameters were identical to those in panel (a) of FIG.~\ref{fig:distribution}. 
    (b) The equilibrium populations of EF-RPMD and Ehrenfest EF-RPMD as a function of temperature. The equilibrium populations we averaged form the trajectory snapshots after the time when population plateaued. 
    The inset demonstrates the RPMD methods slightly deviate from the detailed balance at very low temperatures. 
    (c) Electron transfer rate as a function of temperatures profile for asymmetric diabatic PESs.
	}
\end{figure}    
\end{center}

For asymmetric PESs, the equilibrium electronic population of the molecular level $N_\text{eq}$ depends on the value $\tilde{E}_d$. 
Specifically,  
\begin{equation}
    \label{eqn:neq}
    N_\text{eq} = \frac{k_{1\gets0}}{k_{1\gets0} + k_{1\gets1}} = f(\tilde{E}_d), 
\end{equation}
which is a direct consequence of the detailed balance (Eq.~\ref{eqn:detailed_balance}). 
In FIG.~\ref{fig:detailedbalance}, we plot the steady state population from RPMD at different temperatures with $\tilde{E}_d = -0.3 \hbar \omega $. 
The asymmetric diabatic PESs are shown in FIG.~\ref{fig:detailedbalance} (a). Again, we prepare our initial population on one well and propagate the dynamics over time. 
We then obtain the values for $N_\text{eq}$ by averaging the population over a period of time after the dynamics have reached steady state. 
In FIG.~\ref{fig:detailedbalance}(b), we plot the population $N_\text{eq}$ from RPMD as a function of temperature. 
We notice that RPMD results agree with the prediction from detailed balance almost quantitatively, despite a slight deviation at lower temperatures lower than $0.25 \hbar \omega$. 
Nonetheless, the equilibrium populations can serve as a proof that our RPMD methods obtain the correct detailed balance approximately.

That being said, despite our RPMD methods predicts the correct equilibrium behaviour, the methods overestimate the electron transfer rate when temperature is very low.
Particularly, we note that both EF-RPMD and Ehrenfest EF-RPMD predict a erroneous turn-over trend for the rate illustrated in FIG.~\ref{fig:detailedbalance} (c). 
Such turn-over trend is nether present in numerical results from QME nor the analytical results from the golden rule rate. 
Instead, the rate converges to the tunneling limit which is a constant that independent of temperature. 
These results show that the our averaged friction on centroid mode approach is not very accurate for asymmetric potentials when the temperature is very low. That being said, we note that such deviations are within the same magnitude as the correct rates. We believe that the RPMD methods can still serve as good approximations for the electron dynamics as long as the temperatures are not too low.

\onecolumngrid
\twocolumngrid

\section{\label{sec:conclusion} Conclusion}

We present a method that combines RPMD with electronic friction (denoted EF-RPMD) to study nonadiabatic dynamics at metal surfaces, particularly when nuclear quantum effects play a crucial role. In such a method, the nonadiabatic effects are captured in electronic friction and frictional force, which act on the centroid mode of the ring-polymer molecular dynamics.  
We demonstrate the validity of our method against numerical comparison with QME dynamics. We show that EF-RPMD can not only predict the correct steady state population but also transient dynamics. Furthermore, the EF-RPMD results predict the correct electron transfer rate near metal surface both in the barrier crossing regime and quantum tunneling regime. Given the accuracy and simplicity of the method, we expect that the EF-RPMD can be very useful to study complex chemical processes on metal surface, e.g. chemisorption, heterogeneous catalysis, electrochemistry. 

Looking forward, since the EF model works only in the weak nonadiabatic limit, we expect that the EF-RPMD breaks down when nonadiabatic effects are very strong. Within the electronic friction model, one can incorporate non-Markovian effects for strong nonadiabatic effects. On-going work will address the question of how to incorporate such non-Markovian effects within the RPMD framework.


\begin{acknowledgments}
We acknowledge the support from Westlake University and the National Natural Science Foundation of China.
\end{acknowledgments}


\appendix
\section{\label{app:integrator} Numerical Integrators for EF-RPMD and Ehrenfest EF-RPMD}
In this section, we will go through the technical details of implementing the EF-RPMD and Ehrenfest EF-RPMD methods. The most crucial step for ensuring efficiency is to decompose the integration of the Hamiltonian $H_s^{N}$. 
Directly propagating the dynamics with velocity-verlet algorithm for $H_s^{N}$ can be very inefficient, which requires a very small time step $\Delta t$ to converge the fast oscillating motion of the ring polymer. (see~\citet[p. 473]{Tuckerman_2010}) In addition, without external thermostats, the microcanonical sampling can be very inefficient as well \cite{normalmodelangevinthermostat-C2010}.

These problems can be easily addressed by integrating the equation in normal mode coordinates \cite{Tuckerman_2010}, and adding an external white-noise Langevin thermostat to the ring polymer \cite{normalmodelangevinthermostat-C2010, BAOABisBetter-Z2017}.
Following ref~\cite{BAOABisBetter-Z2017}, we split the total Liouville operator $\mathcal{L}$ corresponding to the total system Hamiltonian $H_s^{N}$, the electronic frictional effect, and the Lagevin thermostats into four terms,
\begin{equation}
    \label{eqn:L_decompose}
	\mathcal{L} = \mathcal{L}_0 + \mathcal{L}_\text{PMF} + \mathcal{L}_\text{ef} + \mathcal{L}_\text{thermostat},
\end{equation}
where $\mathcal{L}_0$, $\mathcal{L}_\text{PMF}$, $\mathcal{L}_\text{ef}$ , and $\mathcal{L}_\text{thermostat}$ denote the Liouville operator for free ring polymer $H_0^{N}$, potential of mean force $\UPMF^N$, the centroid mode electronic friction, and the stochastic Langevin thermostats, respectively. Here, we define $\UPMF^N = E^{N} + V_0^{N}$ with $E^{N}$ and $V_0^{N}$ being introduced in Eq.~\ref{eqn:Hs_path}. Note that the exact form of $\UPMF^{N}$ and the equations of motions correspond to $\mathcal{L}_\text{ef}$ are different for EF-RPMD and Ehrenfest EF-RPMD. Specifically,
\begin{equation}
    \label{eqn:UPMF}
    \UPMF = \left\{
    \begin{array}{ll}
         \sum_{i}^{N} f(E(x_i)) E(x_i) + \frac{1}{2} M x_i^2, & \quad \text{EF-RPMD}  \\
         \sum_{i}^{N} \sigma^{(i)} E(x_i) + \frac{1}{2} M x_i^2, & \quad \text{Ehrenfest EF-RPMD} 
    \end{array}
    \right.
\end{equation}
and $\mathcal{L}_\text{ef}$ corresponds to
\begin{equation}
    \label{eqn:centroideom}
\begin{gathered}
    \left\{
    \begin{array}{ll}
         \dot{\tilde{P}}_0  = -\hbar \gamma_\text{e}^\text{c} \tilde{P}_0 + \delta F^\text{c}, & \quad \text{EF-RPMD} \\
         \dot{\tilde{P}}_0  = \delta F^\text{c}, & \quad \text{Ehrenfest EF-RPMD}
    \end{array}
    \right.
\end{gathered}
\end{equation}
Once again, $\tilde{P}^{(0)}$ represents the centroid ($0$-th) normal mode momentum coordinate, which we will define in the next paragraph.
Lastly, please refer to Eq.(32) in ref.~\cite{normalmodelangevinthermostat-C2010} for the equations of motion corresponding to $\mathcal{L}_\text{thermostat}$

We can diagonalize $H_0^{N}$ with the following orthogonal transformation:
\begin{equation}
	\label{eqn:CartesianToNormalMode}
	\tilde{P}_j = \sum_{i=0}^{N-1} P_i C_{ij} \quad \text{and} \quad
	\tilde{x}_j = \sum_{i=0}^{N-1} x_i C_{ij},
\end{equation}
where the Cartesian coordinates ${x_i, p_i}$ are transformed to normal mode coordinate ${\tilde{x}_j, \tilde{P}_j}$ with orthogonal transformation matrix $C$:
\begin{equation}
	C_{jk} = \left\{
	\begin{aligned}
		 & \sqrt{1/N},                    & j = 0,                     \\
		 & \sqrt{2/N} \cos(\pi j k / N),  & 1 \leq j \leq N / 2 - 1,   \\
		 & \sqrt{1/N} (-1)^{N},           & j = N/2,                   \\
		 & \sqrt{2/N} \sin(\pi j k / N) , & N/2 + 1 \leq j \leq N - 1.
	\end{aligned}
	\right.
\end{equation}
In addition, one can easily inverse the transformation by
\begin{equation}
	\label{eqn:NormalModeToCartesian}
	P_i = \sum_{j=0}^{N-1} C_{ij} \tilde{P}_i \quad \text{and} \quad
	x_i = \sum_{j=0}^{N-1} C_{ij} \tilde{x}_i.
\end{equation}
The transformation diagonalizes the free ring polymer Hamiltonian $H_0^{N}$ in the following form:
\begin{equation}
	\label{eqn:H0_free}
	H_0^{N}(\bm{\tilde{x}}, \bm{\tilde{P}}) = \sum_{k=0}^{N-1} \left( \frac{\tilde{P}_k^2 }{2M} +%
	\frac{1}{2} M \omega_k^2 \tilde{x}_k^2  \right),
\end{equation}
where $\omega_k$ denotes the frequency of the $k$-th normal mode, which is given by
\begin{equation}
	\omega_k = 2 \omega_N \sin(k\pi/N).
\end{equation}
The transformation enables separation between the centroid motion (mode $k=0$) and oscillating motion (mode $k \neq 0$) of the ring polymer. 
With the transformation, one can work out the analytical equations of motion for each normal modes as follows:
\begin{subequations}
	\label{eqn:RPMDFree}
	\begin{gather}
		\dot{\tilde{x}}_0 = \hbar \omega \tilde{P}_0, \\
		\left\{
		\begin{aligned}
			\tilde{P}_j(t) & = \cos{(\omega_j t)} \, \tilde{P}_j(0) - M \omega_j \sin{(\omega_j t)} \, \tilde{x}_j(0)  \\
			\tilde{x}_j(t) & = \frac{1}{M \omega_j} \sin{(\omega_j t)} \, \tilde{P}_j(0) -  \cos{(\omega_j t)} \, \tilde{x}_j(0)
		\end{aligned}
		\right.
	\end{gather}
\end{subequations}
Equations of motion in Eq.\ref{eqn:RPMDFree} corresponds to $\mathcal{L}_0$.
Thus, we have clarified the normal mode coordinates appeared in the main text.

Finally, we present the integrator for the EF-RPMD and Ehrenfest EF-RPMD methods. We adopt a similar "BAOAB" integrator proposed in ref.~\cite{BAOABisBetter-Z2017}, since such scheme can provide numerical stability. In this work, we define $\mathcal{L}_\text{A} \equiv \mathcal{L}_0$, $\mathcal{L}_\text{B} \equiv \mathcal{L}_\text{PMF}$, and $\mathcal{L}_\text{O}\equiv \mathcal{L}_\text{ef} + \mathcal{L}_\text{thermostat}$. Our integrator then reads:
\begin{equation}
    e^{\mathcal{L}\Delta t} \approx%
        e^{\mathcal{L}_\text{B}\Delta t/2} e^{\mathcal{L}_\text{A}\Delta t/2}%
        e^{\mathcal{L}_\text{O}\Delta t}%
        e^{\mathcal{L}_\text{A}\Delta t/2}e^{\mathcal{L}_\text{B}\Delta t/2}%
\end{equation}

\section{\label{app:breakdown} Mean field Ehrenfest perspective of the EF-RPMD and their failure when $\Gamma$ is small.}

In this section, we show that the MDEF approach can be understood from a mean field point of view \cite{EFPRL-D2017, EFREV-D2018}.
In this section, we will also demonstrate that both MDEF and EF-RPMD methods fail when we are in the strong nonadiabatic or non-Markovian regime, i.e., $\Gamma < \hbar\omega$.

To begin, we start with the mean field expansion for the electronic population \cite{EFPRL-D2017, EFREV-D2018},
\begin{equation}
    \label{eqn:expansion}
    \sigma(t; x) = \sigma_\text{eq}(x) + \delta \sigma(t; x)
                 = \sigma_\text{eq}(x) - %
                   \frac{1}{\Gamma} \pdv{\sigma_\text{eq}}{x} \pdv{x}{t}.
\end{equation}
Here, the first term is the instantaneous equilibrium population; the second term is the first order correction due to nuclear motion. With this expansion, we see the second term will be small when $\Gamma > \hbar\omega$. In other words, expansion Eq.~\ref{eqn:expansion} is valid in the Markovian regime.

Next, we argue the Langevin equations of motion Eq.~\ref{eqn:langevin} can be understood using mean field expansion Eq.~\ref{eqn:expansion}.
Through the expansion of $\sigma(t; x)$, and 
\[
    \pdv{\sigma_\text{eq}}{x} = \pdv{f(E(x))}{x} = -\beta \pdv{E}{x} f(E(x)) (1 - f(E(x))) ,
\]
we can derive the equations of motion for the following Ehrenfest dynamics with random force $\delta F(t)$,
\begin{equation}
    \begin{gathered}
    \label{eqn:ehrenfest}
    \dot{x} = \frac{P}{M}, \quad \dot{P} = F_\text{EF} + \delta F(t), \\
    \dot{\sigma} = \frac{\Gamma}{\hbar} (\sigma_\text{eq}(x) - \sigma) = \frac{\Gamma}{\hbar} (f(E(x)) - \sigma).
    \end{gathered}
\end{equation}
Here, $F_\text{EF}$ denotes the Ehrenfest mean force:
\begin{equation}
    \label{eqn:ehrenfest_force}
    F_\text{EF} = -\pdv{E}{x} \sigma(t, x) - \pdv{V_0}{x}.
\end{equation} 
It is straightforward to verify Eq.~\ref{eqn:ehrenfest_force} is equivalent to the potential of mean force in Eq.~\ref{eqn:PMF} plus the frictional force in Eq.~\ref{eqn:langevin}.
Moreover, we add the random force from the Fluctuation-Dissipation theorem (Eq.~\ref{eqn:fluc-diss}) as $\delta F(t)$. 
By this construction, we see Ehrenfest dynamics with random force  $\delta F(t)$ is equivalent with the MDEF Langevin dynamics. And the validity of the derivation depends on the validity of expansion Eq.~\ref{eqn:expansion}.

From our argument, we demonstrate the Langevin dynamics suggest a mean field expansion for $\sigma(t)$. This understanding explains why the MDEF approach is effective for weak nonadiabatic regime, not for the strong nonadiabatic regime. 
Next, we will demonstrate that such limitation for MDEF is also present in its RPMD version, EF-RPMD.

Once again, the conjecture of averaged friction on centroid mode can be rationed with the mean field perspective.
Specifically, the average electronic density of a ring polymer $\sigma^\text{RP}$ can be evaluated as
\begin{equation}
    \sigma^\text{RP}(t; \{x_k\}) = \frac{1}{N} \sum_i^{N} \sigma^{(i)}(t; x_i).
\end{equation}
We then expand the electronic density of the $i$-th replica first order in centroid momentum
\begin{equation}
    \label{eqn:expansion_RP}
    \begin{aligned}
    \sigma(t; x_i) 
        &= \sigma_\text{eq}(x_i) - \frac{1}{\Gamma} \pdv{\sigma^\text{RP}_\text{eq}(t; \{x_k\})}{x_i} \pdv{x_i}{t} , \\
        &=  \sigma_\text{eq}(x_i) -
           \frac{1}{\Gamma}\frac{1}{N} \pdv{\sigma_\text{eq}}{x_i} \pdv{x_i}{\tilde{x}_0} \pdv{\tilde{x}_0}{t} , \\
        &= \sigma_\text{eq}(x_i) - \frac{1}{\Gamma}\pdv{\sigma_\text{eq}}{x_i} \frac{1}{N}\sum_k^{N} \frac{P_k}{M}.
    \end{aligned}
\end{equation}
We have used $\tilde{x}_0 = \frac{1}{\sqrt{N}}\sum_i x_i$ and  $\tilde{P}_0 = \frac{1}{\sqrt{N}}\sum_i P_i$ in the last equality, which is a property of orthogonal transformation matrix $C$. 
With above mean field expansion, we see that the averaged force for the $i$-th bead is
\begin{equation}
    \begin{aligned}
        F_\text{EF}^{(i)} &= -\pdv{E}{x_i} \sigma^{(i)}(t; x_i)  -\pdv{V_0^N}{x_i}, \\
                          &= \bar{F}^{(i)} - \gamma_\text{e}(x_i) \frac{1}{N} \sum_k^{N} \frac{P_k}{M} \\
                          &\approx \bar{F}^{(i)} - \gamma_\text{e}^\text{c} \frac{1}{N} \sum_k^{N} \frac{P_k}{M}
    \end{aligned} \label{eqn:meanF_Ehrenfest_RPMD}
\end{equation}
where $\bar{F}^{(i)}$ in the second line is exactly the Langevin dynamics mean force defined in Eq.\ref{eqn:EF-RPMDmeanF}. 
Finally, to get the simpler averaged friction expression in Eq.~\ref{eqn:centroid_mode_eom}, we replace $\gamma_\text{e}(x_i)$ by the averaged electronic friction $\gamma_\text{e}^\text{c}$. 
Together with the fluctuation-dissipation theorem in Eq.~\ref{eqn:fluc-diss-RP}, we obtain our EF-RPMD, 
\[
    \dot{P}_i = \bar{F}^{(i)} - \gamma_\text{e}^\text{c} \frac{1}{N} \sum_k^{N} \frac{P_k}{M} + \delta F^\text{c},
\]
from our physical intuition.

With the argument above, we see that averaged friction scheme used by our EF-RPMD can be partially explained by the following key points:
1) the molecular orbital densities corresponding to the ring polymer beads reaches instantaneous steady state $\sigma_\text{eq}(x_i)$;
2) the local density fluctuations to $\sigma_\text{eq}(x_i)$ are affected by the collective centroid motion of the ring polymer (Eq.~\ref{eqn:expansion_RP}).

In addition to justifying EF-RPMD, mean field expansion Eq.~\ref{eqn:expansion_RP} also help us the construction of mean field Ehrenfest version of EF-RPMD.
Naively, we can solve the coupled equations of motion of 
\begin{gather}
    \dot{\sigma}^{(i)} = \frac{\Gamma}{\hbar} (f(E(x_i) - \sigma^{(i)}) 
    \label{eqn:RPMDsigmaEOM}, 
\end{gather}
and, 
\[
    \dot{P}_i = F_\text{EF}^{(i)} + \delta F^\text{c},
\]
However, this scheme does not work. This is because the expansion in Eq.~\ref{eqn:expansion_RP} indicates that the electronic density of the $i$-th replica $\dot{\sigma}^{(i)}(t)$ couples with other replicas. In contrast, Eqn.~\ref{eqn:RPMDsigmaEOM} suggests that the bead are is independent.
Therefore, the random force $\delta F^\text{c}$ generated from fluctuation-dissipation theorem Eq.~\ref{eqn:fluc-diss-RP} does not correspond to equations of motion in Eq. \ref{eqn:RPMDsigmaEOM}. Luckily, this problem can be solved by introducing an \textit{ad hoc} force corrections on each bead $F_\text{corr}^{(i)}$. 

To see how these force corrections arise, we first write down the electronic density expansion when there is no inter-bead correlation:
\begin{equation}
    \label{eqn:expansion_RP_independent}
    \begin{aligned}
    \sigma(t; x_i) 
        &= \sigma_\text{eq}(x_i) - \frac{1}{\Gamma} \pdv{\sigma_\text{eq}}{x_i} \pdv{x_i}{t} , \\
        &= \sigma_\text{eq}(x_i) - \frac{1}{\Gamma} \pdv{\sigma_\text{eq}}{x_i} P_i.
    \end{aligned}
\end{equation}
Comparing this equation and expansion Eq~\ref{eqn:expansion_RP}, we see the only difference is the momentum. Specifically, If we correct the $P_i$ into the averaged momentum $\frac{1}{N}\sum_i P_i$, we can get the desired expansion.
Hence, instead of $F_\text{EF}^{(i)} = - \pdv{E}{x} \sigma(t)^{(i)}$, we should instead have 
\begin{equation*}
    - \pdv{E}{x} 
    \left[ \sigma(t)^{(i)} + (P_i - \frac{1}{N} \sum_i P_i)
    \frac{1}{\Gamma} \pdv{\sigma_\text{eq}}{x_i} 
     \right] = F_\text{EF}^{(i)} + F_\text{corr}^{(i)},
\end{equation*}
which simplifies into 
\begin{equation}
    F_\text{corr}^{(i)} = \gamma_\text{e}(x_i) \Delta p_i,%
                          \quad \Delta p_i = p_i - \frac{1}{N} \sum_i p_i, \label{eqn:Fcorr} %
\end{equation}
Overall, Equations of motion for the Ehrenfest dynamics that is equivalent to EF-RPMD:
\begin{equation}
    \label{eqn:centroid_mode_eom_Ehrenfest}
    \dot{P}_i  = F_\text{EF}^{(i)} + F_\text{corr}^{(i)} + \delta F^\text{c}.
\end{equation}
Again, the controid mode random force $\delta F^\text{c}$ is be generated by Eq.~\ref{eqn:fluc-diss-RP}.

As both EF-RPMD and its Ehrenfest analogue are derived from the mean field treatment of the electronic DOF ($d^{\dagger}d$), these methods are only valid when there is a clear separation of timescales between the nuclear motion and electronic relaxation. 
Specifically, the electronic relaxation need to be much faster than the nuclear timescales.
For the chemisorption model, transient dynamics predicted by our EF-RPMD will fail when $\Gamma < \hbar \omega$. 

\begin{figure}[htbp]
    \centering
    \includegraphics[width=0.49\textwidth]{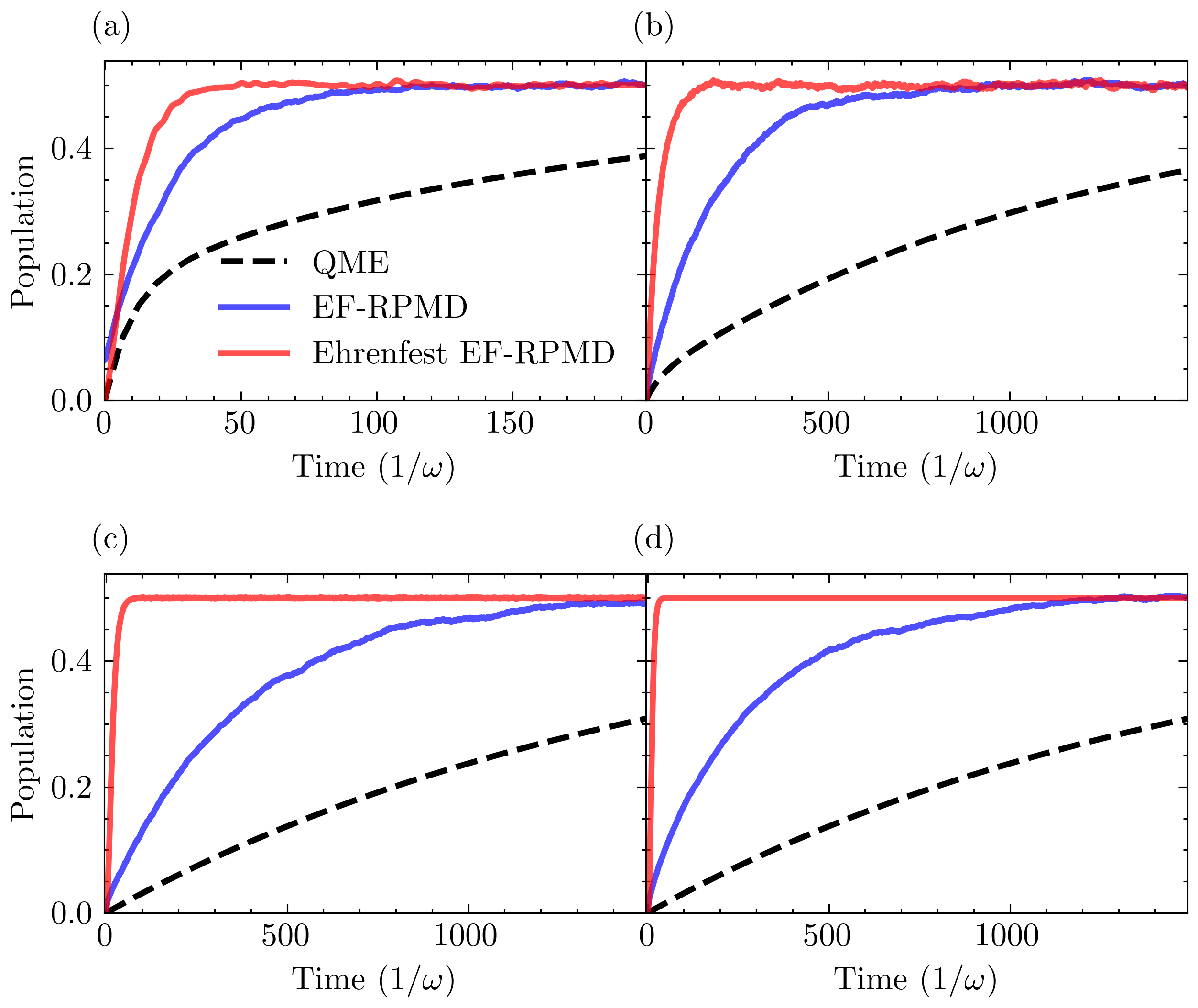}
    \caption{\label{fig:population-nonMark} Electronic population of the single molecular level as a function of temperatures in strong non-adiabatic regime.
    All parameters remain identical to those in FIG.~\ref{fig:population}, except for $\Gamma = 0.001 = 0.333 \hbar \omega$.
    This demonstrates the two RPMD methods are not applicable for strong non-adiabatic systems.}
\end{figure}

FIG.~\ref{fig:population-nonMark} demonstrates the breakdown of both EF-RPMD and its Ehrenfest counterpart when $\Gamma = 0.333\hbar \omega < \hbar \omega$.
Note that our electronic friction based methods predicts way faster electronic relaxation dynamics than the true quantum dynamics. 
Such result manifests the importance of the memory effect, which is ignored in our model.
Note also that, the transient dynamics predicted by EF-RPMD and mean field dynamics does not agree anymore,
suggesting first order the mean field expansion of electronic density becomes invalid when $\Gamma$ gets larger.



\section{\label{app:rate} Thermally Averaged Golden-Rule Rates: the Barrier-Crossing and Tunneling Limits}
In ref.~\cite{QCMEderive-D2015}, we have shown when $k_\text{B} T \gg \text{ZPE}$, the classical limit of the forward $k_{1\gets0}$ and backward $k_{0 \gets1 }$ electron transfer rate (i.e. Marcus rates) are
\begin{subequations}
    \begin{align}
        k_{1 \gets 0} &= \int \dd{\epsilon} \Gamma f(\epsilon) 
        \frac{e^{-(E_\text{r} - \epsilon + \tilde{E}_d )^2/4 E_\text{r} k_\text{B}T}}{\sqrt{4 \pi E_\text{r} k_\text{B}T}}, \\
         k_{0 \gets 1} &= \int \dd{\epsilon} \Gamma (1 - f(\epsilon)) 
        \frac{e^{-(E_\text{r} + \epsilon - \tilde{E}_d )^2/4 E_\text{r} k_\text{B}T}}{\sqrt{4 \pi E_\text{r} k_\text{B}T}}.
    \end{align}
\end{subequations}    
These equations predict that when $k_\text{B} T \to 0$, both rates will vanish, leading to the prediction of no electron transfer. 
In contrast, we have numerically demonstrated in Section~\ref{subsec:ET} that the electron transfer rates at low temperatures converge to a non-vanishing constant value from both QME and our RPMD methods. We denote such limits as the tunneling tunneling regime.
To rationalize such temperature independent nature of ET rate at low temperatures, we will derive the rate analytically in the following.

First, we will use a quantum description for the nuclear DOFs in the system Hamiltonian. Specifically, coordinates $x$ and $p$ in Eq.~\ref{eqn:Hs} can be quantized by introducing the ladder operators $a$ and $a^{\dagger}$:
\begin{equation}
    H_\text{s} = E_d  d^{\dagger} d + 
    g (a + a^{\dagger}) \numb{d} + \hbar \omega(a^{\dagger} a + \frac{1}{2}).
\end{equation}
To proceed, we specify the following polaron transform associated with the electron-phonon interaction and apply such transformation to the total Hamiltonian $H$.
For arbitrary operator $O$, we denote the polaron transform as $\tilde{O}\equiv \hat{U}_d O \hat{U}_d^{-1}$, where the transformation matrix $\hat{U}_d$ is defined as
\begin{equation}
    \hat{U}_d = 
    \exp[\bar{\lambda} d^{\dagger} d (a^{\dagger} - a)] ,
\end{equation}
with $\bar{\lambda} = g / \hbar \omega$ denoting the nuclear shift related to interaction $(a + a^{\dagger}) \numb{d}$. The polaron transformed total Hamiltonian $\tilde{H} = \hat{U}_d H \hat{U}_d^{-1}$ reads
\begin{equation}
    \label{eqn:H-polaron}
    \begin{aligned}
        \tilde{H} &= \hbar \omega(\numb{a} + \frac{1}{2}) + 
        \tilde{E}_d \numb{d} + 
        \sum_k (\epsilon_k - \mu) \numb{c_k} +  \tilde{H}_\text{c},\\
        \tilde{H}_\text{c} &= \sum_k V_k \left(c_k^{\dagger} d e^{-\bar{\lambda} (a^{\dagger} - a)} + d^{\dagger} c_k e^{\bar{\lambda} (a^{\dagger} - a)} \right),
    \end{aligned}
\end{equation}
and we will denote the polaron transformed interaction Hamiltonian as $\tilde{H}_\text{c}$ (second line of Eq.~\ref{eqn:H-polaron}).
After the transform, note that the orbital energy $E_d$ is re-normalized to $\tilde{E}_d$.

Second, we will evaluate the thermally averaged electron transfer rates. Take $k_{1\gets0}$ as an example, we can calculate the averaged golden-rule rate by summing over all possible transitions between all coupled nuclear-electronic states weighted by a Boltzmann factor for the initial phonon-state and a Fermi function factor for the bath electronic state. Specifically, 
\begin{equation}
    \label{eqn:deri-rate01}
    \begin{aligned}
        k_{1 \gets 0} = \, &\frac{2\pi}{\hbar} \sum_k f(\epsilon_k) \sum_\nu P_\nu^\text{Boltz} \sum_{\nu^{\prime}} \delta (\epsilon_k - \tilde{E}_d - (\nu^{\prime} - \nu) \hbar \omega) \\
        &\abs{\mel{\nu^{\prime}, n_k = 0, n=1}{\tilde{H}_\text{c}}{\nu, n_k = 1, n=0}}^2,
    \end{aligned}
\end{equation}
where the Boltzmann factor of the initial nuclear distribution can be written as
\begin{equation*}
    P_\nu^\text{Boltz} = \frac{e^{-\nu \hbar \omega / k_\text{B} T}}{\sum_\nu e^{-\nu \hbar \omega / k_\text{B} T}}.
\end{equation*}
Eq.~\ref{eqn:deri-rate01} can be readily simplified by noting that only the $d^{\dagger} c_k$ term in $\tilde{H}_\text{c}$ will survive. Thus, 
\begin{equation*}
    \begin{gathered}
        \abs{\mel{\nu^{\prime}, n_k = 0, n=1}{\tilde{H}_\text{c}}{\nu, n_k = 1, n=0}}^2 = \\
        \abs{\mel{\nu^{\prime}}{V_k e^{\bar{\lambda}(a^{\dagger} - a)}}{\nu}}^2 = \abs{V_k}^2 \text{FC}_{\nu^{\prime}, \nu}^2, 
    \end{gathered}
\end{equation*}
where the squared term simplifies into a product of $\abs{V_k}^2$ and a squared Frank-Condon factor denoted as $\text{FC}_{\nu^{\prime}, \nu}^2$. 
Such factor can be readily evaluated by \cite{FC-Koch2004}:
\begin{equation}
    \text{FC}_{\nu^{\prime}, \nu} = \frac{p!}{Q!} \bar{\lambda}^{Q - p} e^{-\frac{\bar{\lambda}^2}{2}} L_{Q-p}^{p} (\bar{\lambda}^2) 
    [\text{sgn}(\nu^{\prime} - \nu)]^{\nu - \nu^{\prime}},
\end{equation}
where $p=\min(\nu, \nu^{\prime})$, $Q = \max(\nu, \nu^{\prime})$, and function $L_{Q-p}^{p} (x)$ denotes the generalized Laguerre polynomial. 
With such simplification, as well as the wide-band approximation (Eq.~\ref{eqn:WBA}),  Eq.~\ref{eqn:deri-rate01} can be further simplified as
\begin{equation}
    \label{eqn:rate-forward}
    \begin{aligned}
        k_{1 \gets 0} = \frac{\Gamma}{\hbar}  
        \sum_\nu P_{\nu}^\text{Boltz} 
        \sum_{\nu^{\prime}} f(\tilde{E}_d + (\nu^{\prime} - \nu) \hbar \omega) \text{FC}_{\nu^{\prime}, \nu}^2.
    \end{aligned}
\end{equation}
In a similar manner, we can derive the backward rate:
\begin{equation}
    \label{eqn:rate-backward}
    \begin{aligned}
        k_{0 \gets 1} = \frac{\Gamma}{\hbar}  
        \sum_\nu P_{\nu}^\text{Boltz} 
        \sum_{\nu^{\prime}} (1 - f(\tilde{E}_d - (\nu^{\prime} - \nu) \hbar \omega)) \text{FC}_{\nu^{\prime}, \nu}^2.
    \end{aligned}
\end{equation}
It is straightforward to verify that the detailed balance condition, 
$k_{1\gets0} = e^{\beta \tilde{E}_d} k_{0\gets1}$, holds true
with Eq.~\ref{eqn:rate-forward}-\ref{eqn:rate-backward}.

Lastly, we will evaluate $k_{1 \gets 0}$ in the low temperature limit, $\hbar \omega \gg k_\text{B} T$.
In such condition, only the vibrational ground state $\nu=0$ for the initial state is significant in the Boltzmann term, thus, 
\begin{equation}
    \label{eqn:rate-lowT}
    \begin{aligned}
        \eval{k_{1 \gets 0}}_{\hbar \omega \gg k_\text{B} T} 
        \approx \frac{\Gamma}{\hbar}   
        \sum_{\nu^{\prime}} f(\tilde{E}_d + \nu^{\prime}  \hbar \omega) \text{FC}_{\nu^{\prime}, 0}^2.
    \end{aligned}
\end{equation}
Furthermore, we could approximate the Fermi function using the following Heaviside step function when temperature is low.
\begin{equation}
    \eval{f(x)}_{\hbar\omega \gg k_\text{B}T} \approx (1 - \Theta(x)), \quad 
    \Theta(x) = \left\{
        \begin{array}{r r}
             0,   &  x < 0,\\
             0.5, &  x = 0,\\
             1,   &  x > 0,
        \end{array} \right.
\end{equation}
1). If $\tilde{E}_d \ge 0$, all the Fermi function terms with $\nu^{\prime} \ge 1$ will vanish: 
\begin{equation}
    \label{eqn:rate-lowT_non-neg}
    \begin{aligned}
        \eval{k_{1 \gets 0}}_{\hbar \omega \gg k_\text{B} T, \tilde{E}_d \ge 0} 
        \approx \frac{\Gamma}{\hbar} f(\tilde{E}_d ) \text{FC}_{0, 0}^2 
        \approx \frac{\Gamma}{\hbar} (1 - \Theta(\tilde{E}_d )) e^{-\tilde{\lambda}^2}.
    \end{aligned}
\end{equation}
2). If $\tilde{E}_d < 0$, we need to sum over all non-vanishing Fermi function factors: 
\begin{equation}
    \label{eqn:rate-lowT-neg}
    \begin{gathered}
        \eval{k_{1 \gets 0}}_{\hbar \omega \gg k_\text{B} T, \tilde{E}_d < 0} 
        \approx 
        \frac{\Gamma}{\hbar}   
        \sum_{\nu^{\prime}} \left[1 - \Theta\left(\nu^{\prime} + \frac{\tilde{E}_d}{\hbar\omega} \right) \right] \text{FC}_{\nu^{\prime}, 0}^2.
    \end{gathered}
\end{equation}
In conclusion, we have obtained the analytical tunneling limits for electronic transfer rates in Eq.~\ref{eqn:rate-lowT_non-neg} and~\ref{eqn:rate-lowT-neg}. 
These limits for rates are temperature-independent constants, which supports the numerical trend presented in the main text (FIG.~\ref{fig:rate}).

\bibliography{aipsamp}

\end{document}